\documentclass[12pt]{article}
\usepackage{graphicx}
\usepackage{color}
\usepackage{colortbl}
\usepackage{rotating}
\usepackage{amsmath} 
\usepackage{ifthen} 
\usepackage{multirow}
\usepackage{subfigure}
\usepackage{rotating}
\usepackage{placeins}
\usepackage{morefloats}
\usepackage{amssymb}
\usepackage{amsfonts}
\usepackage{upgreek} 
\usepackage{xspace}
\usepackage{hyperref}    
\usepackage[all]{hypcap} 
\usepackage{amsmath} 
\usepackage{amssymb}
\usepackage{amsfonts}
\usepackage{bm}
\usepackage{upgreek} 
\usepackage{microtype}
\usepackage{lineno}  
\usepackage{xspace} 

\newboolean{uprightparticles}
\setboolean{uprightparticles}{false} 

\def\lhcb {\mbox{LHCb}\xspace}

\def\babar  {\mbox{BaBar}\xspace}
\def\belle  {\mbox{Belle}\xspace}
\def\cdf    {\mbox{CDF}\xspace}

\ifthenelse{\boolean{uprightparticles}}%
{

 \def\Ppi         {\ensuremath{\uppi}\xspace}

 \def\PDelta      {\ensuremath{\Delta}\xspace}                 
 \def\PXi      {\ensuremath{\Xi}\xspace}                 
 \def\PLambda      {\ensuremath{\Lambda}\xspace}                 
 \def\PSigma      {\ensuremath{\Sigma}\xspace}                 
 \def\POmega      {\ensuremath{\Omega}\xspace}                 
 \def\PUpsilon      {\ensuremath{\Upsilon}\xspace}                 
 
 \def\PD      {\ensuremath{\mathrm{D}}\xspace}

 \def\PK      {\ensuremath{\mathrm{K}}\xspace}

 \def\Pb      {\ensuremath{\mathrm{b}}\xspace}

 \def\Pi      {\ensuremath{\mathrm{i}}\xspace}

}
{

 \def\Ppi         {\ensuremath{\pi}\xspace}

 \mathchardef\PDelta="7101
 \mathchardef\PXi="7104
 \mathchardef\PLambda="7103
 \mathchardef\PSigma="7106
 \mathchardef\POmega="710A
 \mathchardef\PUpsilon="7107

 \def\PD      {\ensuremath{D}\xspace}

 \def\PK      {\ensuremath{K}\xspace}

 \def\Pb      {\ensuremath{b}\xspace}

 \def\Pi      {\ensuremath{i}\xspace}

}

\def\bquark    {\ensuremath{\Pb}\xspace}

\def\pion  {\ensuremath{\Ppi}\xspace}

\def\pip   {\ensuremath{\pion^+}\xspace}
\def\pim   {\ensuremath{\pion^-}\xspace}
\def\pipi  {\ensuremath{\pion^+\pion^-}\xspace}

\def\kaon  {\ensuremath{\PK}\xspace}
  \def\Kbar  {\kern 0.2em\overline{\kern -0.2em \PK}{}\xspace}

\def\Kz    {\ensuremath{\kaon^0}\xspace}
\def\Kzb   {\ensuremath{\Kbar^0}\xspace}
\def\KzKzb {\ensuremath{\Kz \kern -0.16em \Kzb}\xspace}
\def\Kp    {\ensuremath{\kaon^+}\xspace}
\def\Km    {\ensuremath{\kaon^-}\xspace}

\def\KpKm  {\ensuremath{\Kp \kern -0.16em \Km}\xspace}

  \def\Dbar    {\kern 0.2em\overline{\kern -0.2em \PD}{}\xspace}
\def\D       {\ensuremath{\PD}\xspace}

\def\Dz      {\ensuremath{\PD^0}\xspace}
\def\Dzb     {\ensuremath{\Dbar^0}\xspace}
\def\DzDzb   {\ensuremath{\Dz {\kern -0.16em \Dzb}}\xspace}
\def\DztoDzb {\ensuremath{\Dz -{\kern -0.16em \Dzb}}\xspace}
\def\Dp      {\ensuremath{\D^+}\xspace}
\def\Dm      {\ensuremath{\D^-}\xspace}

\def\DpDm    {\ensuremath{\Dp {\kern -0.16em \Dm}}\xspace}
\def\Dstar   {\ensuremath{\PD^*}\xspace}

\def\Dstarpm {\ensuremath{\PD^{*\pm}}\xspace}

\def\CP                {\ensuremath{C\!P}\xspace}

\newcommand{\decay}[2]{\ensuremath{#1\!\to #2}\xspace}         

\def\to                 {\ensuremath{\rightarrow}\xspace}

\def\xprime     {\ensuremath{x^{\prime}}\xspace}
\def\xprimesq   {\ensuremath{x^{\prime 2}}\xspace}
\def\yprime     {\ensuremath{y^{\prime}}\xspace}

\def\agamma     {\ensuremath{A_{\Gamma}}\xspace}

\def\kk         {\ensuremath{\PK\PK}\xspace}

\def\dzkk       {\decay{\Dz}{\Kp\Km}}
\def\dzpipi     {\decay{\Dz}{\pip\pim}}
\def\dzbkk      {\decay{\Dzb}{\Kp\Km}}
\def\dzbpipi    {\decay{\Dzb}{\pip\pim}}
\def\dkpicf     {\decay{\Dz}{\Km\pip}}
\def\dkpiws     {\decay{\Dz}{\Kp\pim}}

\def\dstarpmDpipm   {\decay{\Dstarpm}{\Dz\Ppi^{\pm}}}

\def\order{{\ensuremath{\cal O}}\xspace}
\newcommand{\chisq}{\ensuremath{\chi^2}\xspace}

\newcommand\pubnumber{CKM2014}
\newcommand\pubdate{\today}

\def\manchester{School of Physics and Astronomy,\\ The University of
  Manchester, Manchester, UK}
\def\support{\footnote{On Behalf of the LHCb Collaboration}}

\def\Title#1{\begin{center} {\Large #1 } \end{center}}
\def\Author#1{\begin{center}{ \sc #1} \end{center}}
\def\Address#1{\begin{center}{ \it #1} \end{center}}

\newcommand\pubblock{\rightline{\begin{tabular}{l} \pubnumber\\
         \pubdate  \end{tabular}}}
\newenvironment{Abstract}{\begin{quotation}  }{\end{quotation}}
\newenvironment{Presented}{\begin{quotation} \begin{center} 
             PRESENTED AT\end{center}\bigskip 
      \begin{center}\begin{large}}{\end{large}\end{center} \end{quotation}}

\def\beq{\begin{equation}}
\def\eeq#1{\label{#1}\end{equation}}
\def\eeqn{\end{equation}}

\def\beqa{\begin{eqnarray}}
\def\eeqa#1{\label{#1}\end{eqnarray}}
\def\eeqan{\end{eqnarray}}

\let\bar=\overbar

\def\D{{\cal D}}

\def\Dslash{\not{\hbox{\kern-4pt $D$}}}
\def\dslash{\not{\hbox{\kern-2pt $\del$}}}

\def\msb{{\bar{\ssstyle M \kern -1pt S}}}

\begin{document}
\begin{titlepage}
\pubblock

\vfill
\Title{D-mixing and indirect \CP violation measurements at LHCb }
\vfill
\Author{ Silvia Borghi\support}
\Address{\manchester}
\vfill
\begin{Abstract}
The \lhcb experiment collected during run I the world's
largest sample of charmed hadrons.  This sample is used to search for
\CP violation in charm and for the measurements of \Dz~mixing parameters.  The
measurement of the \DztoDzb mixing parameters and the search
for indirect \CP-violation in two-body charm decays at LHCb experiment
are presented.

\end{Abstract}
\vfill
\begin{Presented}
Presented at the 8th International Workshop on the CKM Unitarity 
            Triangle (CKM 2014)
\\ Vienna, Austria, September 8-12, 2014 
\end{Presented}
\vfill
\end{titlepage}
\def\thefootnote{\fnsymbol{footnote}}
\setcounter{footnote}{0}

\section{Introduction}
The charm sector is a promising field to probe for the effects of
physics  beyond the Standard Model (SM). Flavour mixing in the charm 
sector is now well established~\cite{hfag}.   In the SM~\cite{theorysm,theorysm2},
the \CP violation in charm transitions is expected to be small, with
asymmetries up to few $\mathcal{O}(10^{-3})$, while
it can be enhanced by contribution from New Physics~\cite{theoryNP}.

The LHCb experiment is dedicated to the study of  $b$ and $c$ flavour
physics.  The abundance of charm particles produced in LHC offers an
unprecedented opportunity for high precision measurements in the charm
sector, including measurements of  \CP violation and \DztoDzb
mixing.

The results of search for indirect \CP violation and the  measurements
of the mixing parameters in two body hadronic \Dz~charm decays  are
presented here.

\section{Mixing and \CP violation with \dkpiws decays}

The flavour mixing occurs because the mass eigenstates 
($|\mbox{D} _{1,2} \rangle$) are linear combinations of the flavour
eigenstates and they can be written as linear combinations of the
flavour eigenstates $|\mbox{D}_{1,2} \rangle=p|\Dz \rangle\pm{}q|\Dzb
\rangle$, with complex  coefficients $p$ and $q$ which satisfy
$|p|^2+|q|^2=1$.  The mixing parameters are defined as $x\equiv
(m_1-m_2)/ \Gamma $ and  $y\equiv (\Gamma_1-\Gamma_2)/(2\Gamma)$,
where $m_1$, $m_2$, $\Gamma_1$ and  $\Gamma_2$ are the masses and the
decay widths for $\mbox{D}_{1}$ and $\mbox{D}_{2}$, respectively,  and
$\Gamma=(\Gamma_1+\Gamma_2)/2$.  The phase convention is chosen such
that $\CP |\Dz \rangle=-|\Dzb\rangle$.

The first evidence of \DztoDzb oscillation was reported in 2007 by
\babar~\cite{babar07} and \belle\cite{belle07}. Now, the mixing in the
charm sector is well established with the first  observation by a
single measurement with greater than 5 standard deviation significance
at the \lhcb experiment \cite{lhcbmixing1}, confirmed by
\cdf~\cite{cdf2013} and \belle~\cite{belle2014}.

At the \lhcb experiment, the charm mixing parameters are
determined by the  decay-time-dependent ratio of \dkpiws (called wrong
sign, WS) to \dkpicf (called right sign, RS) decay rates. The RS decay
rate is dominated by Cabibbo favoured (CF) amplitude. The WS rate arises
from the interfering amplitudes of the doubly Cabibbo-suppressed decay (DCS) 
and the CF decay following \DztoDzb oscillation.
Assuming no \CP violation and small mixing parameters ($|x|$ and $|y|$ $\ll
1$), this ratio is:
$$ R(t) \approx R_D + \sqrt{R_D} \yprime \frac{t}{\tau} +\frac{\xprimesq+y^2}{4}
\left( \frac{t}{\tau} \right)^2
$$ where $\xprime=x \cos \delta + y \sin \delta$, $\yprime=y \cos \delta - x
\sin \delta$, $R_D$ is the ratio of suppressed-to-favoured decay
rates, $\delta$ is the strong phase difference between the DCS decays 
and the CF decays
$\left( \mathcal{A}\left(\dkpiws \right)/\mathcal{A}\left(\dkpicf \right)=- \sqrt{R_D} e^{-i
  \delta}\right)$.

One can search for \CP violation in the charm sector by comparing
the time-dependent ratios evaluated separately for the two flavours
(\Dz~and \Dzb).  A difference in the $R_D$ parameter between \Dz~and \Dzb
would be a sign of direct \CP violation. While a difference in $\xprimesq$
and $\yprime$ parameters would imply an indirect \CP violation contribution
($|q/p|\neq 1$ or $\phi=arg(q/p)\neq 0$). The data are fit considering three
scenarios:
1) assuming \CP conservation 2) allowing only indirect \CP
contribution; 2) allowing both direct and indirect \CP contributions.

The full data sample with a total integrated luminosity of 3~fb$^{-1}$
is used to perform these measurements.  The analysis is performed on
\dstarpmDpipm decays to allow the determination of the flavour of the
neutral \PD meson at production.

\begin{figure}[!bt]
\centering
\includegraphics[width=\textwidth]{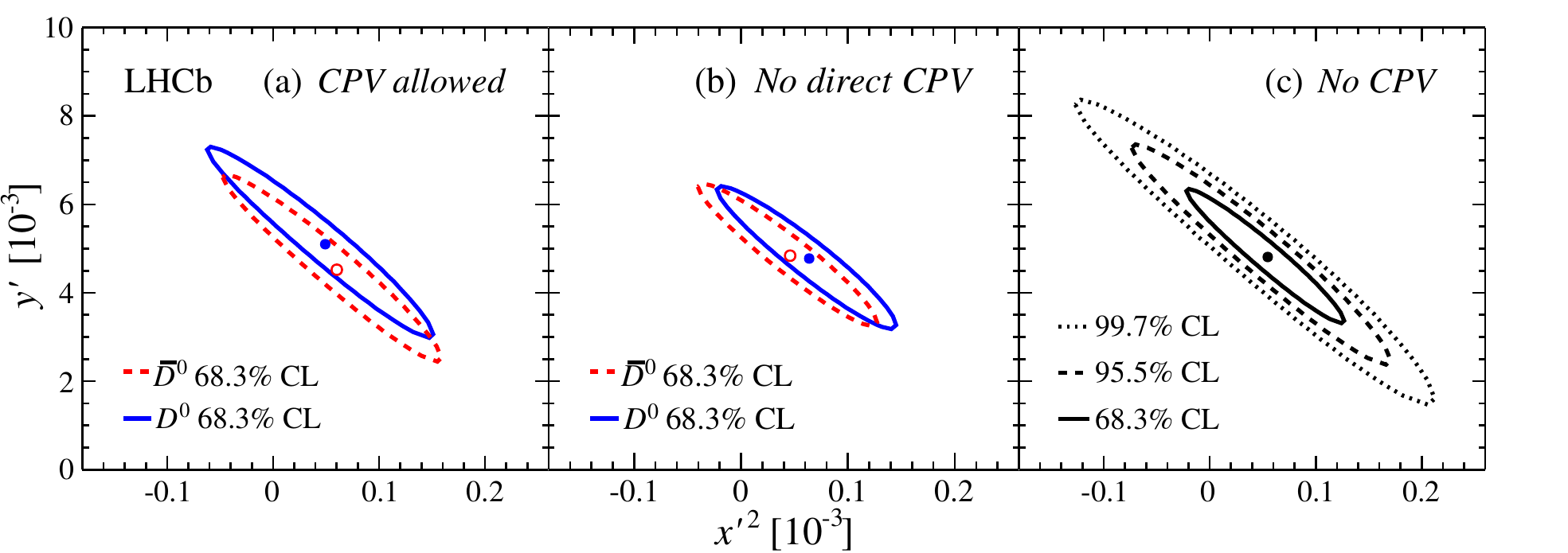}
\caption{Two-dimensional confidence regions in the ($\xprimesq$, $\yprime$)
  plane obtained a) without any restriction on \CP violation, b) assuming no
  direct \CP violation and c) assuming \CP conservation. The solid (dashed)
  curves in a) and b) indicate the contours of the mixing parameters
  associated with \Dz~(\Dzb) decays.  The solid, dashed and dotted
  curves in c) indicate the contours of \CP-averaged mixing parameters
  at $68.3\%$, $95.5\%$ and $99.7\%$ C.L.. The best-fit value is shown
  with a point~\cite{lhcbmixing2}.}
\label{fig3}
\end{figure}

In the scenario 1) the mixing parameters are measured to be $R_D=
(3.568\pm 0.058\pm 0.033)\cdot 10^{-3}$, $\yprime = (4.8 \pm 0.8\pm
0.5)\cdot 10^{-3}$ and $\xprimesq =( 5.5 \pm 4.2 \pm 2.6)\cdot 10^{-5}$
where the first uncertainty is statistical and the second systematic.
In the scenario 2) and 3), the magnitude of $| q/p |$ is
constructed. The magnitude of $q/p$ is determined to be $0.75 < | q/p |
< 1.24$ and $0.67 <| q/p |<1.52$ at $68.3\%$ and $95.5\%$ C.L.,
respectively~\cite{lhcbmixing2}.  The results of the mixing parameters measurements and
of the confidence regions are shown in Fig.~\ref{fig3} for the three
scenarios. They are compatible with \CP conservation and
provide the most stringent bounds on the parameter $| q/p |$ from a
single experiment.


\section{Indirect \CP violation in 2-body \Dz~decays to \CP eingenstates}
A measurement of the indirect \CP violation in \Dz~mixing can be
performed in the study of two-body hadronic charm decays to \CP
eigenstates (\dzkk or \dzpipi).   It can be evaluated by the asymmetry in 
the effective lifetimes ($\tau$) of flavour-tagged decays and it can be
expressed by the following equation with the assumption of negligible
direct \CP-violation contribution:
$$
\agamma=\frac{\tau  (\Dzb \rightarrow f)-\tau (\Dz \rightarrow f)}
                 {\tau(\Dzb \rightarrow f)+\tau (\Dz \rightarrow f)}
\approx \frac{1}{2}\: A_m\: y\:  \cos \phi - x\: \sin \phi 
$$ 
where $A_m$ is defined by  $\mid
q/p \mid^{\pm 2} \approx 1 \pm A_m$.  A measurement of \agamma~differing significantly
from zero is a  manifestation of indirect \CP violation as it requires
a non-zero value for $A_m$ or $\phi$.

The measurement of \agamma~at \lhcb is performed using 1~fb$^{-1}$ of data from
a sample of $pp$ collisions at a centre-of-mass energy of 7 TeV
collected in 2011.  
The flavour at production is again determined using neutral \PD mesons
from \dstarpmDpipm decays.
The main selection is applied at the trigger level
on the momentum, PID and impact parameter (IP) of the \Dz~daughters.
 The trigger event selection
causes a bias of the proper-time distribution. Thus, an acceptance
correction is needed for the evaluation of the effective lifetime.
The acceptance is determined using a data driven method, the so-called
swimming algorithm~\cite{swimming3,agammafirst}.

\begin{figure}[!tb]
\centering \includegraphics[width=0.47\textwidth]{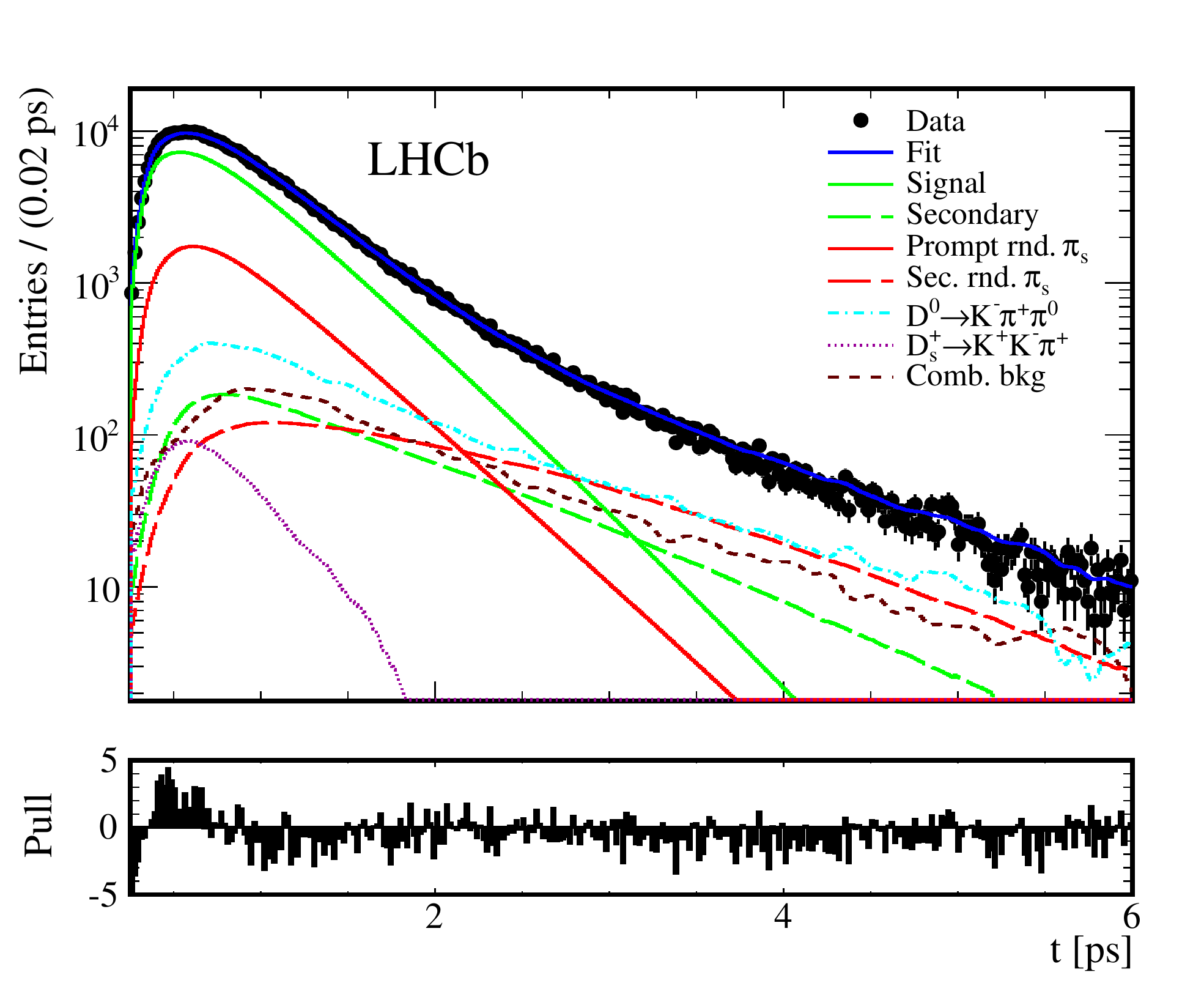}
\includegraphics[width=0.47\textwidth]{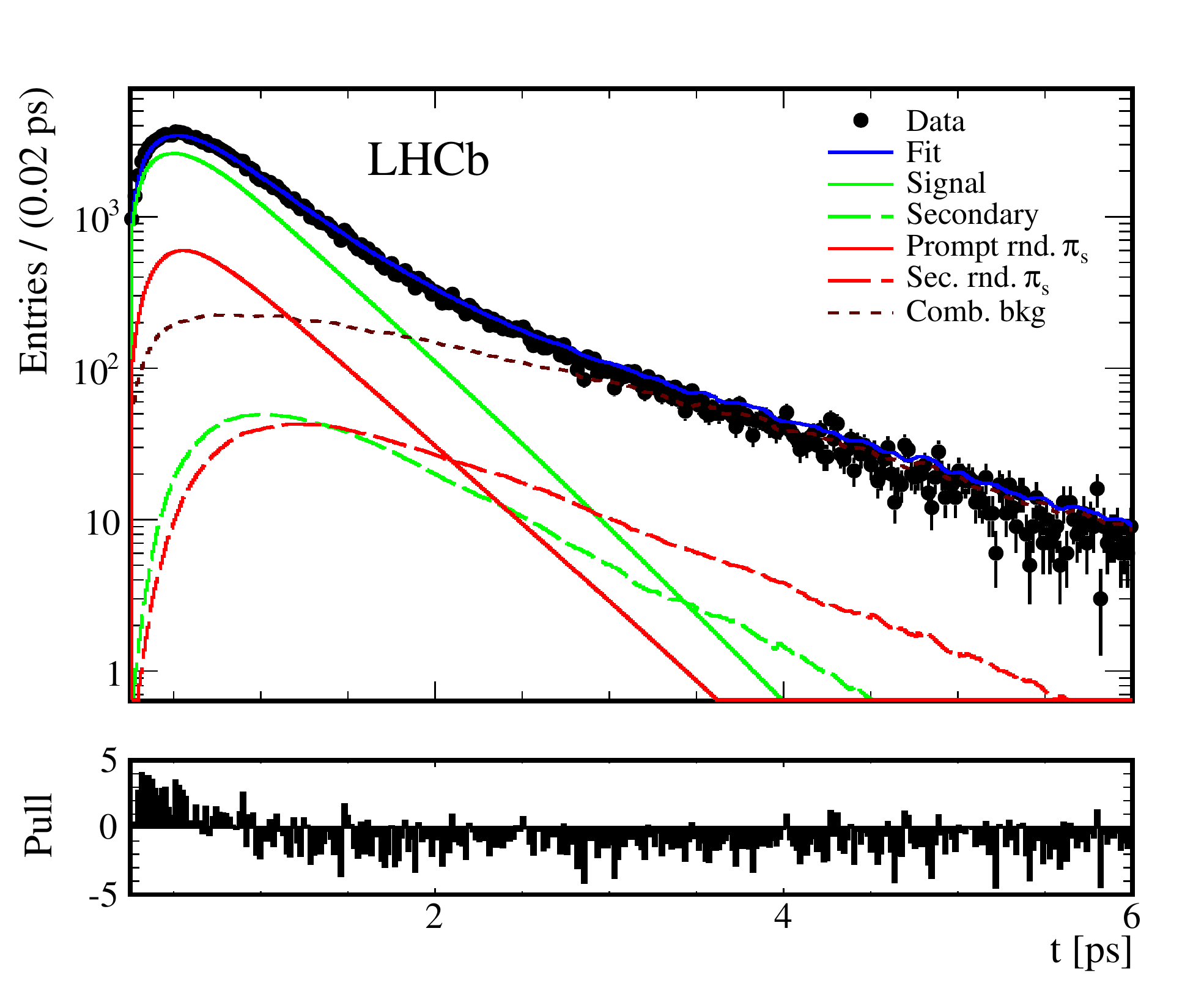}
\includegraphics[width=0.47\textwidth]{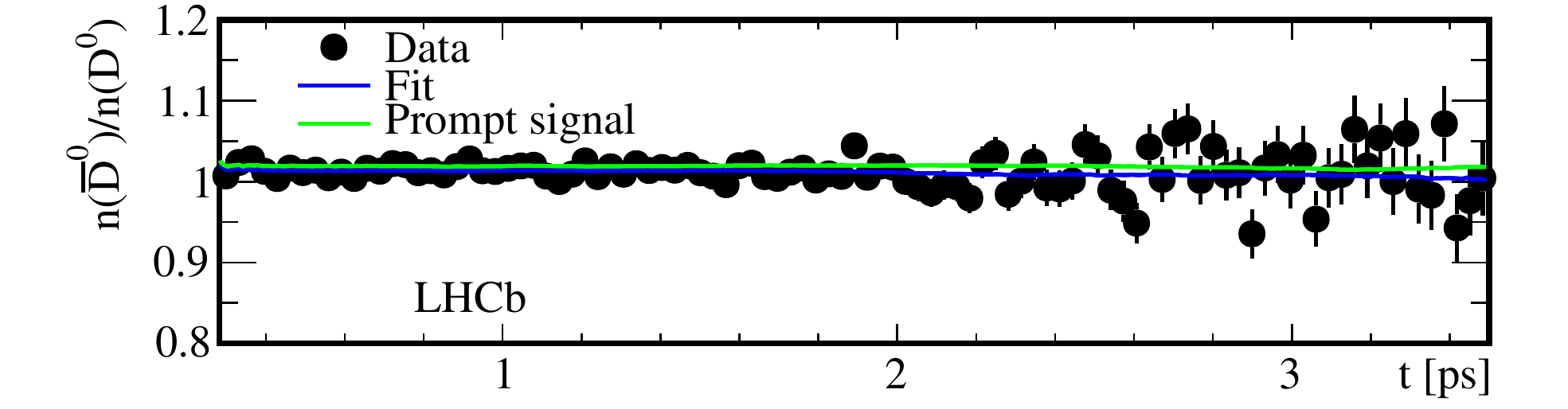}
\includegraphics[width=0.47\textwidth]{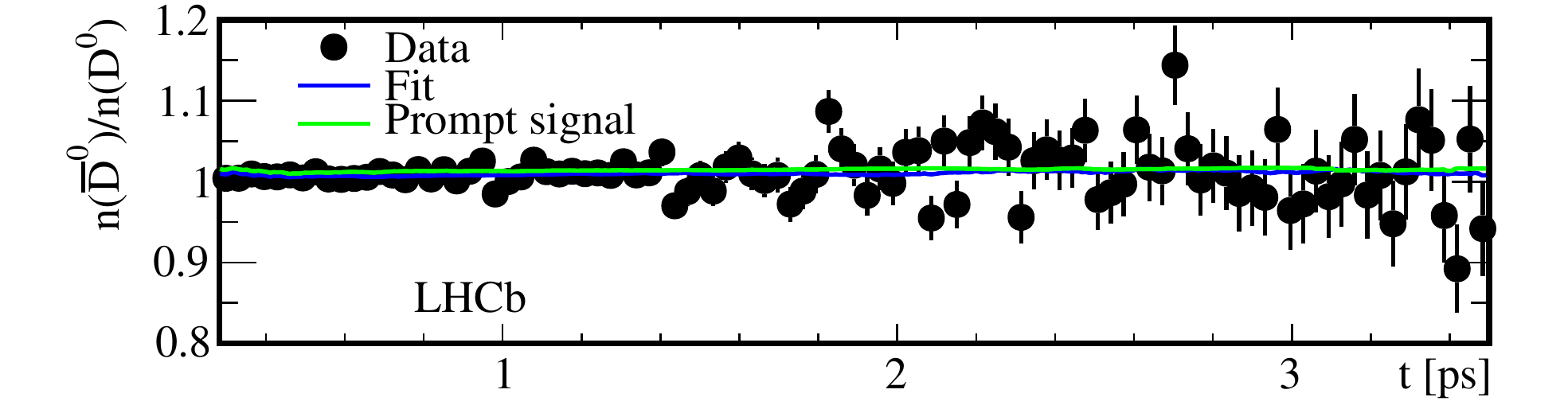}
\caption{Lifetime fit projection of  \dzbkk decays (on the top left)
  and of \dzbpipi decays (on top right) and corresponding pull plot,
  for one data set. The ratio of \Dzb to \Dz~data and fit model for
  decays to \kk (on the left bottom) and \pipi (on the right bottom)
  for all data, respectively~\cite{agamma}.}
\label{figagamma}
\end{figure}

The fit to determine the effective lifetime is performed independently
for each flavour tag and each decay mode.   The signal yield are 
extracted from  simultaneous unbinned
likelihood  fits of the \Dz~invariant mass and of the difference
between \Dstar and \Dz masses, $\Delta$m,
to distinguish the different background contributions.  Charm mesons 
produced in \bquark-hadron decays, secondary charm, have larger impact
parameter with respect to the primary vertex than the prompt
candidates as a secondary $D$ does not usually  point back to the
primary vertex.  This additional background  is subtracted in
in a simultaneous fit of the proper-time and  
$\mbox{ln} \left( \chisq_{IP} \right)$ distributions. 
The $\chisq_{IP}$ is defined as the difference in $\chisq$ of a given
primary interaction vertex reconstructed with and without the
considered particle.  Fig.~\ref{figagamma} shows an example of the
proper-time projection for \dzkk and \dzpipi decays for one data set. 

The analysis method is validated on a control sample of CF \dkpicf
decays, where the lifetime asymmetry is determined to
be consistent with zero in accordance with the  expectation.  The
resulting values of \agamma~for the two final states~\cite{agamma} are:
\begin{eqnarray*}
\agamma^{KK} & = & (-0.35\pm0.62\pm 0.12)\cdot 10^{-3}\nonumber\\
\agamma^{\pi\pi} & = & (0.33\pm1.06\pm0.14)\cdot 10^{-3}\\
\end{eqnarray*}
where the first uncertainty is statistical and the second systematic.
Both results are consistent with zero, showing no evidence for 
indirect \CP violation. They are consistent with and more precise than 
previous determinations from other experiments \cite{hfag}.
Amongst the several sources of
systematics considered, the main ones are due to the decay-time
acceptance correction and due to the background description. 

\section{Conclusion}
The mixing in charm sector is now well established, while searches
for indirect \CP violation yield results consistent with CP conservation. 
Further measurements are  ongoing at LHCb using the data set collected 
during Run I and others will follow with the upcoming Run 2, allowing to 
explore the charm sector with unprecedented precisions. Later, the \lhcb 
Upgrade is expected to collect 50 fb$^{-1}$ of integrated luminosity, 
allowing to reach precisions down to $0.5 \cdot 10^{-3}$ for \agamma and 
$\order(10^{-5},10^{-4})$ for $(\xprimesq,\yprime)$.



\end{document}